\documentclass[a4paper,11pt]{article}
\pdfoutput=1

\usepackage{jheppub} 

\newcommand{\la}{\langle}
\newcommand{\ra}{\rangle}
\newcommand{\ep}{\epsilon}
\newcommand{\s}{\sigma}

\newcommand{\mN}{\mathcal{N}}
\newcommand{\mA}{\mathcal{A}}
\newcommand{\mM}{\mathcal{M}}
\newcommand{\mP}{\mathcal{P}}
\newcommand{\tmA}{\widetilde{\mathcal{A}}}

\newcommand{\sumn}{\sigma\in S_n/(\mathbb{Z}_n\times\mathbb{Z}_2)}
\newcommand{\sump}{\sigma\in S_5/(\mathbb{Z}_5\times\mathbb{Z}_2)}
\newcommand{\sumh}{\sigma\in S_6/(\mathbb{Z}_6\times\mathbb{Z}_2)}

\newcommand{\prescription}{{double-copy construction}}
\newcommand{\spacesection}{\vskip 0.5in}

\title{\boldmath Virtual Color-Kinematics Duality:\\$6$-pt $1$-Loop MHV Amplitudes}

\author{Ellis Ye Yuan}
\affiliation{Perimeter Institute for Theoretical Physics,\\31 Caroline St.~N., Waterloo, Ontario N2L 2Y5, Canada}
\affiliation{Department of Physics and Astronomy, University of Waterloo,\\200 University Avenue W., Waterloo, Ontario N2L 3G1, Canada}
\emailAdd{yyuan@perimeterinstitute.ca}

\abstract{We study $1$-loop MHV amplitudes in $\mN=4$ super Yang-Mills theory and in $\mN=8$ supergravity. For Yang-Mills we find that the simple form for the full amplitude presented by Del Duca, Dixon and Maltoni naturally leads to one that has physical residues on all compact contours. After expanding the simple form in terms of standard scalar integrals, we introduce redundancies under certain symmetry considerations to impose the color-kinematics duality of Bern, Carrasco and Johansson (BCJ). For five particles we directly find the results of Carrasco and Johansson as well as a new compact form for the supergravity amplitude. For six particles we find that all kinematic dual Jacobi identities are encapsulated in a single functional equation relating the expansion coefficients. By the BCJ \prescription{} we obtain a formula for the corresponding $\mN=8$ supergravity amplitude. Quite surprisingly, all physical information becomes independent of the expansion coefficients modulo the functional equation. In other words, there is no need to solve the functional equation at all. This is quite welcome as the functional equation we find, using our restricted set of redundancies, actually has no solutions. For this reason we call these results virtual color-kinematics duality. We end with speculations about the meaning of an interesting global vs.~local feature of the functional equation and the situation at higher points.}

\begin{document}
\maketitle
\flushbottom


\section{Introduction and Main Results}
\label{sec:introduction}

Originally proposed by Bern, Carrasco and Johansson \citep{Bern2008,Bern2010a}, the remarkable discovery of color-kinematics duality provides a powerful tool for exploring gravity amplitudes by starting from Yang-Mills amplitudes and ``squaring'' the kinematic numerators (double-copy construction). This technique has led to a new way of thinking about tree and loop amplitudes. At tree level the validity of this construction has been proved \citep{Bern2010}, and the existence of duality-respecting kinematic numerators has been explicitly shown for all numbers of external particles \citep{Bern2008,Bjerrum-Bohr2011,Mafra2011,Bjerrum-Bohr2012}. At loop level, arguments from unitary cuts and soft limit strongly suggest that the double-copy construction should hold as long as the duality is achieved \citep{Bern2010a,Bern2010,Oxburgh2012}. This together with the validity of a duality-respecting formula has be confirmed in various cases for $4$ points up to $4$ loops and for $5$ points up to $2$ loops \citep{Bern2010a,Carrasco2012,Carrasco2011,Bern2011,Boucher-Veronneau2011,Naculich2012,Bern2012,Bern2012a}, the situation at $4$ points $5$ loops is currently under exploration \cite{Bern2012b}, and the self-dual sector in Yang-Mills has also been studied \citep{Monteiro2011}. A natural next step is to study $1$-loop MHV amplitudes with $n\geq6$ as well as the explicit connection between the maximally supersymmetric Yang-Mills theory and gravity here. The purpose of this paper is to give some first steps in this direction.

In the 90's, Del Duca, Dixon and Maltoni showed that one-loop amplitudes in Yang-Mills can be written as~\footnote{Here we use a notation $\mA^{(L)}$ (or $A^{(L)}$) to indicate the level of the amplitude, where $L$ is the number of loops, and $L=0$ refers to tree-level amplitudes. The curly letter denotes the full amplitudes and the ordinary letter denotes the partial amplitudes. The tilde sign indicates that the formula is written without helicity factor or SUSY delta functions. Yang-Mills amplitudes are denoted by the letter $\mA$ (or $A$) and gravity amplitudes by $\mM$.}~\citep{DelDuca2000b}
\begin{equation}\label{eq:result1}
  \mA^{(1)}_n=g^{n}_{\rm YM} \sum_{\sumn}f_{\s_1\s_2\ldots\s_n}
  A^{(1)}_n(\s_1\s_2\ldots\s_n)
\end{equation}
where $f_{\s_1\s_2\ldots\s_n}$ is a combination of structure constants $f_{abc}$ contracted in a necklace-like form and  $A^{(1)}_n(\s_1\s_2\ldots\s_n)$ is the so-called leading color partial amplitude of the more standard color-decomposition~\citep{Dixon1996}.

In \citep{Arkani-Hamed2010,Arkani-Hamed2011}, a formula for $A^{(1)}_n(\s_1\s_2\ldots\s_n)$ in terms of special integrals which gives physical results on all compact contours respecting the color-ordering was presented. We write it as
\begin{equation}\label{eq:result1a}
  A^{(1)}_n(\s_1\s_2\ldots\s_n) = \frac{\delta^4(p_1+\cdots +p_n)\delta^8(\lambda_1\widetilde{\eta}_1+\cdots +\lambda_n\widetilde{\eta}_n)}{\la\s_1\s_2\ldots\s_n\ra}\mP_n(\s_1\s_2\ldots\s_n)
\end{equation}
where $\la\s_1\s_2\ldots\s_n\ra$ is a shorthand notation for the standard Parke-Taylor denominator \citep{Parke1986}. In the rest of the paper we will not write the (super) momentum conservation delta functions and the coupling constant in any amplitude in order not to clutter equations.

In this paper, we show that a simple combination of \eqref{eq:result1} and \eqref{eq:result1a}
\begin{equation}\label{eq:newYMconjecture}
  \tmA^{(1)}_n=\sum_{\sumn}\frac{f_{\s_1\s_2\ldots\s_n}}{\la\s_1\s_2\ldots\s_n\ra}
  \mP_n(\s_1\s_2\ldots\s_n),
\end{equation}
gives a formula that matches physical residues on all compact contours.

We propose to use $\mP_n(\s_1\s_2\ldots\s_n)$ as a basis of integrals not only in Yang-Mills but also for gravity amplitudes. The intention is to seek for a clearer observation of the relations between the two theories. In brief, these are totally ``planar'' objects enjoying cyclic and reflection symmetry as well as simple residues on all contours, and we propose to call them MHV polygons.

By expanding the MHV polygons onto scalar loop integrals and studying the color-kinematics duality, we manage to recover Carrasco and Johansson's result for $5$-pt case in $\mN=8$ supergravity \citep{Carrasco2012}, and further transform it back to our formulation, which is as simple and compact as that in Yang-Mills theory
\begin{equation}\label{eq:result3}
  \mM^{(1)}_5=\sum_{\sump}\frac{1}{\la\s_1\s_2\s_3\s_4\s_5\ra}\frac{[\s_1\s_2\s_3\s_4\s_5]}{\epsilon(\s_1\s_2\s_3\s_4)}\mP_5(\s_1\s_2\s_3\s_4\s_5).
\end{equation}
Here $[\s_1\s_2\s_3\s_4\s_5]$ denotes the Parke-Taylor denominator of an $\overline{{\rm MHV}}$ tree amplitude while $\epsilon(\s_1\s_2\s_3\s_4)$ is the standard fully anti-symmetric tensor contracted with four vectors.

Then as we go on to $6$-pt case, we start to observe a new phenomenon. The formula for the gravity amplitude can still be expressed nicely as
\begin{equation}\label{eq:result4}
  \mM^{(1)}_6=\sum_{\sumh}\frac{\gamma(\s_1\s_2\s_3\s_4\s_5\s_6)}{\la\s_1\s_2\s_3\s_4\s_5\s_6\ra^2}\mP_6(\s_1\s_2\s_3\s_4\s_5\s_6),
\end{equation}
with the coefficients $\gamma$ required to satisfy a set of constraints, each of which relates two $\gamma$'s that differ only by a transposition of two adjacent labels; e.g.~for the standard ordering, we have
\begin{equation}\label{eq:result4a}
  \begin{split}
    \frac{\gamma(123456)}{\la123456\ra\ep(123456)}+\frac{\gamma(213456)}{\la213456\ra\ep(213456)}&\\
    +\frac{[12][34][56]}{\la12\ra\la34\ra\la56\ra}\frac{s_{12}s_{34}s_{45}s_{56}\la3|4+5|6][3|4+5|6\ra}{\ep(123456)\ep(213456)\ep(3456)}&=0,
  \end{split}
\end{equation}
where $\epsilon(abcdef)=\la ab\ra[bc]\la cd\ra[de]\la ef\ra[fa]-[ab]\la bc\ra[cd]\la de\ra[ef]\la fa\ra$. For convenience, we refer to these constraints as a single \emph{functional} constraint, in the sense that it depends on a choice of permutation of the labels. Quite surprisingly, all physical information in $\mM^{(1)}_6$ becomes $\gamma$ independent. More explicitly, once $\mP_6(\s_1\s_2\s_3\s_4\s_5\s_6)$ is expanded, on the physical $\mathbb{R}^4$ contour, in terms of only scalar box integrals the corresponding coefficients become $\gamma$ independent after using the functional equation. This fact is actually quite welcome as the functional equation has no solutions!

The functional equation \eqref{eq:result4a} is found by first expanding the MHV polygons in the Yang-Mills amplitude in terms of scalar integrals. We then add some particular set of redundancies to impose the color-kinematics duality conditions. It turns out that all conditions boil down to the single functional equation \eqref{eq:result4a}. Using \prescription{} we obtain the corresponding gravity amplitude. Just as for five particles, it is possible to transform it back to our formulation leading to the formula presented above.

The fact that the functional equation has no solutions implies that the numerators we found in Yang-Mills which in principle satisfy the color-kinematics duality do not actually exist. This should {\it not} come as a surprise as our ansatz for the redundancies was {\it not} the most general one. What is surprising is that even with a ``virtual" solution to the color-kinematics duality conditions one can still square the numerators to get the corresponding gravity amplitude.

The paper is organized as follows. In Section \ref{sec:polygon} we introduce MHV polygons and summarize their general properties, after which we prove the formula for $n$-pt $1$-loop MHV amplitudes in super Yang-Mills constructed with these polygons. Then in Section \ref{sec:5ptcase} we analyze the $5$-pt case to give a taste of the relation between this new formulation and the color-kinematics duality in the context discussed by Carrasco and Johansson. Afterwards, in Section \ref{sec:6ptcase} we focus on the simplest case where new phenomena would arise, where we analyze the color structure, obtain the condition on which color-kinematics duality is satisfied, and explain the ``inconsistency'' that appears. In Section \ref{sec:6ptsugra} we go on to test the corresponding conjectured supergravity amplitude, where we provide evidences from quadruple cuts. The details of this analysis as well as a short introduction to color-kinematics duality are summarized in the appendices. In the end, we list out some possible future explorations along this line.

\spacesection

\section{MHV Polygons and Their Properties}
\label{sec:polygon}

As has been stated in the introduction, we use chiral tensor integrals that possess unit leading singularities to build $1$-loop MHV amplitudes \citep{Arkani-Hamed2010,Arkani-Hamed2011}, in order to bring all physically important information into the coefficients attached to them. Specifically within the context of our discussion, we call these objects $n$-pt MHV polygons (or MHV $n$-gons), and symbolically denote them as $\mP_n(\s_1\s_2\ldots\s_n)$. They are functions of the configuration of external particles (both the ordering and the kinematics data), and can be defined in two equivalent ways. One is that an $n$-pt MHV polygon shares the same configuration of leading singularities with its corresponding $1$-loop MHV Yang-Mills partial amplitude, but with the values of all non-zero leading singularities normalized. The other definition, as can be seen in \eqref{eq:result1a}, is by the following equation
\begin{equation}\label{eq:polygondefinition}
  \mP_n(\s_1\s_2\ldots\s_n)
  =\frac{\widetilde{A}^{(1)}_n(\s_1\s_2\ldots\s_n)}{\widetilde{A}^{(0)}_n(\s_1\s_2\ldots\s_n)}.
\end{equation}
The equivalence of the two definitions is due to the fact that for any quadruple cut on the $1$-loop MHV Yang-Mills amplitude, the associated leading singularity (whenever it is non-zero) is always the corresponding tree-level amplitude \citep{Cachazo2008}.

By definition, these objects should be invariant under both cyclic permutation and reflection of the sequence of the particle labels
\begin{align}
  \label{eq:polygoncycl}\mP_n(\s_1\s_2\ldots\s_{n-1}\s_n)&=\mP_n(\s_2\s_3\ldots\s_n\s_1),\\
  \label{eq:polygonrefl}\mP_n(\s_1\s_2\ldots\s_{n-1}\s_n)&=\mP_n(\s_n\s_{n-1}\ldots\s_2\s_1).
\end{align}
That is to say, for $n$ particles, the length of the set $\{\mP_n\}$ is $(n-1)!/2$, and each MHV polygon is a planar object with a fixed ordering of particle labels.

It is always possible to expand MHV polygons as a linear combination of scalar pentagon integrals and scalar box integrals, and the general method to obtain this expansion is worked out in \citep{Arkani-Hamed2011}. Here we will only list out the $5$-pt and $6$-pt cases, which are needed in subsequent analysis.

For any MHV pentagon (e.g.~$\mP_5(12345)$), there is one unique reduction formula
\begin{equation}\label{eq:pentagonreduction}
  \begin{split}
    \mP_5(12345)&=\frac{s_{12}s_{23}s_{34}s_{45}s_{51}}{\ep(1234)}I_5(1|2|3|4|5)\\
    &+\sum_{\mathbb{Z}_5}\frac{\la34\ra[45]\la51\ra[13]s_{12}s_{23}}{\ep(1234)}I_4(1|2|3|45),
  \end{split}
\end{equation}
where the summation in the second line is performed over cyclic permutations of the sequence $(12345)$; $I_n$ denotes the ordinary scalar loop integrals with $n$ loop propagators, and the vertical bars in the arguments separate external legs into groups that connect to different vertices on the loop. We also use the notation $s_{ab}=(p_a+p_b)^2$, and the $\ep$ symbol is defined by
\begin{equation}\label{eq:epsilon4def}
  \begin{split}
    \ep(abcd)&=4i\ep_{\mu\nu\rho\sigma}p^\mu_ap^\nu_bp^\rho_cp^\sigma_d\\
    &=\la ab\ra[bc]\la cd\ra[da]-[ab]\la bc\ra[cd]\la da\ra,
  \end{split}
\end{equation}
and so it is completely antisymmetric in its arguments.

Redundancies begin to appear in the reduction of MHV hexagons (e.g.~$\mP_6(123456)$). From the method as given in \citep{Arkani-Hamed2011}, by fixing label $1$ and $6$, we can obtain one specific formula
\begin{equation}\label{eq:hexagonreduction1}
  \begin{split}
    \mP_6(123456)
    &=-\sum\frac{s_{34}s_{45}s_{56}(3|4+5|6)^2}{\ep(3456)}I_5(12|3|4|5|6)\\
    &\quad-\sum\frac{\la12\ra[23]\la34\ra[41](1|2+3|4)^2}{\ep(1234)}I_4(1|23|4|56)\\
    &\quad+\sum\frac{s_{56}\ep(123456)(5|3+4|6)^2}{\ep(3456)\ep(5612)}I_4(12|34|5|6)\\
    &\quad+\sum s_{45}s_{56}\frac{\la34\ra[46][3|4+5|6\ra}{\ep(3456)}I_4(123|4|5|6)\\
    &\quad+\sum s_{34}s_{45}\frac{[35]\la56\ra\la3|4+5|6]}{\ep(3456)}I_4(612|3|4|5).
  \end{split}
\end{equation}
In the above expression, each summation is performed over cyclic permutations of even steps. And we use the following abbreviations
\begin{align}
  \label{eq:parenthesis}(a|b+c|d)^2&=\la a|b+c|d][a|b+c|d\ra,\\
  \label{eq:epsilon6def}\ep(abcdef)&=\la ab\ra[bc]\la cd\ra[de]\la ef\ra[fa]-[ab]\la bc\ra[cd]\la de\ra[ef]\la fa\ra.
\end{align}
In \eqref{eq:parenthesis} the square is just a notation to keep track of the dimension. We can observe that $\ep(abcdef)$ is invariant under cyclic permutation of even steps, and acquires a minus sign under cyclic permutation of odd steps.

Note that, \eqref{eq:hexagonreduction1} is manifestly invariant only under cyclic permutations of the labels by even steps, and it consists of only half of all the pentagon and box integrals with the correct ordering. Although cyclic permutations of odd steps also leads to a valid formula (which is by itself also manifestly invariant under cyclic permutations of even steps), their equivalence is only guaranteed by non-trivial identities among the pentagon and box integrals. Let us denote these two reductions as $\mathcal{H}(123456)$ and $\mathcal{H}(234561)$ respectively. With the purpose of exploring color-kinematics duality later on in the context of loop integrals, we should exhaust the entire loop integral basis. And so we need to express the MHV hexagon as a linear combination of the two reductions
\begin{equation}\label{eq:fullhexreduction}
  \mP_6(123456)=\alpha(123456)\mathcal{H}(123456)+\beta(123456)\mathcal{H}(234561),
\end{equation}
under the constraint
\begin{equation}\label{eq:alphaconstraint}
  \alpha(123456)+\beta(123456)=1.
\end{equation}
Here the argument in the parenthesis is pure labeling and identical up to cyclic permutations of even steps. If we regard them as functions of particles' kinematics data, the functions should respect this symmetry. But notice that by now there is no definition for e.g.~$\alpha(234561)$ and $\beta(234561)$, so that we can further identify
\begin{equation}
  \alpha(234561)=\beta(123456),
\end{equation}
and require that as functions, $\alpha(234561)$ and $\alpha(123456)$ are related by permutations. Then both the $\alpha$ functional parameter and the reduction formula \eqref{eq:fullhexreduction} manifestly enjoy the full cyclic invariance.

\spacesection

\section{$1$-Loop MHV Amplitudes in $\mN=4$ Super Yang-Mills}
\label{sec:yangmills}

We first go on to show that given the MHV polygons defined in the previous section, the formula \eqref{eq:newYMconjecture}
\begin{equation*}
  \tmA^{(1)}_n=\sum_{\sumn}\frac{f_{\s_1\s_2\ldots\s_n}}{\la\s_1\s_2\ldots\s_n\ra}
  \mP_n(\s_1\s_2\ldots\s_n)
\end{equation*}
matches correct physical residues on all contours. In order to do this, we need to use a special formula for the tree-level amplitude \citep{DelDuca2000c}, which can be easily proved by Britto-Cachazo-Feng-Witten (BCFW) method \citep{Britto2005d,Britto2005e}
\begin{equation}\label{eq:chainformula}
    \tmA^{(0)}_{n}=\sum_{\sigma\in S_{n-2}}\frac{f_{1\sigma_2,\sigma_3,\ldots,\sigma_{n-2},\sigma_{n-1}n}}{\langle1\sigma_2\sigma_3\cdots\sigma_{n-1}n\rangle},
\end{equation}
where we take the abbreviation
\begin{equation}
  f_{1\sigma_2,\sigma_3,\ldots,\sigma_{n-2},\sigma_{n-1}n}=f_{1\sigma_2a_2}f_{a_2\sigma_3a_3}\cdots f_{a_{n-2}\sigma_{n-1}n},
\end{equation}
and the permutation $\sigma_i$ is taken over the label set $\{2,3,\ldots,n-1\}$. This can be visualized as the summation of chain-like diagrams where we fix the two ends of the ``chain'' (in the given formula we fix $1$ and $n$) and fully permute all the intermediate vertices.

Then we go on to evaluate the factorization of $1$-loop full amplitude in the quadruple cuts in $\mN=4$ super Yang-Mills. Since in \eqref{eq:chainformula} each color factor is accompanied by a kinematic factor which has exactly the form of an MHV tree-level partial amplitude, we can start by evaluating the quadruple cuts on partial amplitude. As has been mentioned previously, in MHV super Yang-Mills amplitudes, whenever the factorization is non-trivial, it always has the form
\begin{equation}\label{eq:factorizationYM}
  A^{(1)}_n\longrightarrow\prod^4A^{(0)}=A^{(0)}_n\cdot\text{Det}(|J|),
\end{equation}
where $\text{Det}(|J|)$ is exactly identical to the Jacobi determinant arising from cutting loop propagators and is to be exactly canceled by that factor under quadruple cut \citep{Cachazo2008}. So what is left to be matched from the ansatz side is purely the Parke-Taylor form.

Now switch to the full amplitude, where the non-trivial factorizations only come from two types of quadruple cuts. The virtue of the formula \eqref{eq:chainformula} is that, due to the freedom in picking up any two external particles as the two ends of the chain, if we always choose in the factorized amplitudes those ``external particles'' from the cut propagators, then as a result the color factors will just glue together to form the maximal loop (in the sense of color diagrams). And when combining the results on the corresponding kinematic factors that mimic partial amplitudes, we conclude the non-vanishing factorizations are always
\begin{equation}\label{eq:factorizationChain}
  \tmA^{(1)}_n\longrightarrow\prod^4\tmA^{(0)}=\text{Det}(|J|)\cdot\sum_{\{\s^{(1)}\}}\sum_{\{\s^{(2)}\}}\sum_{\{\s^{(3)}\}}\sum_{\{\s^{(4)}\}}\frac{f_{\{\s^{(1)}\}\{\s^{(2)}\}\{\s^{(3)}\}\{\s^{(4)}\}}}{\la\{\s^{(1)}\}\{\s^{(2)}\}\{\s^{(3)}\}\{\s^{(4)}\}\ra},
\end{equation}
where the summation is performed over the group of external particles attached to each factorized amplitude respectively. This is exactly what we would get if the formula sums over all non-equivalent MHV polygons and the coefficient in front of each MHV polygon has the form as shown in \eqref{eq:newYMconjecture}. So we conclude that with the MHV polygons defined in \eqref{eq:polygondefinition} as the fundamental building block, the formula for $1$-loop $n$-pt MHV super Yang-Mills full amplitude is
\begin{equation}\label{eq:newYMformula}
  \tmA^{(1)}_n=\sum_{\sumn}\frac{f_{\s_1\s_2\ldots\s_n}}{\la\s_1\s_2\ldots\s_n\ra}
  \mP_n(\s_1\s_2\ldots\s_n).
\end{equation}

\spacesection

\section{From Yang-Mills to Gravity: $5$-pt $1$-Loop Amplitudes}
\label{sec:5ptcase}

In order to provide an example of how our formulation may work between Yang-Mills and gravity, in this section we focus on the $5$-pt case, showing that our formula
\begin{equation}\label{eq:5pt1loopYM}
  \tmA^{(1)}_5=\sum_{\sump}\frac{f_{\s_1\s_2\s_3\s_4\s_5}}{\la\s_1\s_2\s_3\s_4\s_5\ra}\mP_5(\s_1\s_2\s_3\s_4\s_5)
\end{equation}
naturally gives rise to color-kinematics duality in its expansion onto scalar loop integrals, and further obtaining the corresponding gravity amplitude as a compact expression also built purely upon MHV pentagons.

In the reduction formula for the MHV pentagons \eqref{eq:pentagonreduction}, it is easy to observe that each $I_5$ receives a unique contribution from its corresponding $\mP_5$, while the coefficient in front of each $I_4$ would receive contributions from two $\mP_5$'s. To make the structure that appears in later discussion more apparent, we define
\begin{equation}\label{eq:Qdefinition}
  Q(\s_1\s_2\s_3\s_4\s_5)=\frac{[\s_1\s_2\s_3\s_4\s_5]}{\ep(\s_1\s_2\s_3\s_4)},
\end{equation}
where similar to the abbreviation for angle brackets, we have
\begin{equation}
    [\s_1\s_2\cdots\s_n]=[\s_1\s_2][\s_2\s_3]\cdots[\s_n\s_1].
\end{equation}
The function $Q$ is totally symmetric in any cyclic permutation of the labels and acquires a minus sign under reflection~\footnote{It is interesting that this $Q$ also played an important role in the analysis of \cite{Carrasco2012}} (This is desirable since $\langle\s_1\s_2\s_3\s_4\s_5\rangle Q(\s_1\s_2\s_3\s_4\s_5)$ is the coefficient of the pentagon integral, and we would prefer it to have the same symmetry with the corresponding basis element). Then the coefficient of one specific $I_4$, e.g.~$I_4(1|2|3|45)$ is
\begin{equation}
  \begin{split}
    f_{12345}&\frac{[12][13][23][45]}{\la45\ra\epsilon(1234)}-f_{12354}\frac{[12][13][23][54]}{\la54\ra\epsilon(1235)}\\
    =&\frac{1}{s_{45}}f_{a1b}f_{b2c}f_{c3d}f_{dea}f_{e45}\left[Q(12345)-Q(12354)\right].
  \end{split}
\end{equation}
So in the expansion the formula is again purely in terms of another functional coefficient $Q$
\begin{equation}\label{eq:5ptCJresult}
  \begin{split}
    \tmA^{(1)}_5&=\sum f_{\s_1\s_2\s_3,\s_4\s_5}\left[Q(\s_1\s_2\s_3\s_4\s_5)-Q(\s_1\s_2\s_3\s_5\s_4)\right]\frac{I_4(\s_1|\s_2|\s_3|\s_4\s_5)}{s_{\s_4\s_5}}\\
    &\quad+\sum f_{\s_1\s_2\s_3\s_4\s_5}Q(\s_1\s_2\s_3\s_4\s_5)I_5(\s_1|\s_2|\s_3|\s_4|\s_5),
  \end{split}
\end{equation}
where we use the abbreviation
\begin{equation}\label{eq:abbref}
  f_{\s_1\s_2\s_3,\s_4\s_5}=f_{a\s_1b}f_{b\s_2c}f_{c\s_3d}f_{dea}f_{e\s_4\s_5},
\end{equation}
and the summations are performed over all nonequivalent permutations of the labels respectively. Very nicely, at this point we may observe that this is exactly Carrasco and Johansson's result obtained in \citep{Carrasco2012} from an ansatz that respects color-kinematics duality~\footnote{For a quick review of the color-kinematics duality, please refer to Appendix \ref{app:CKDuality}.}, where the basis are $\{I_4/s,I_5\}$.

A more interesting implication is that the MHV pentagons in the new formulation naturally encodes the color-kinematics duality in an implicit way. In fact, we can modify the expansion of $\mP_5$ \eqref{eq:pentagonreduction} into a different form, e.g.
\begin{equation}\label{eq:pentagonexpansion2}
  \begin{split}
    \mP_5(12345)
    &=\la12345\ra\bigg\{Q(12345)I_5(1|2|3|4|5)+\sum_{\mathbb{Z}_5}\frac{Q(12345)-Q(12354)}{s_{45}}I_4(1|2|3|45)\bigg\}.
  \end{split}
\end{equation}
Hence we observe that, the MHV polygons may have a special reduction formula where all the coefficients of lower-order scalar loop integrals can be generated solely by the one that corresponds to its unique highest-order scalar loop integral. For this particular $5$-pt case, this structure comes right from the unique reduction formula of MHV pentagons.

To further appreciate the power of this $Q$ functional coefficient, we go on to $1$-loop $5$-pt gravity amplitude. Since the structure of the expansion onto loop integrals satisfies the full color-kinematics duality, by the \prescription{} originally proposed by Bern, Carrasco and Johansson \citep{Bern2010a}, we can immediately substitute the color factor by another copy of the kinematic factor in each term in the the expansion, and the resulted formula is expected to be the correct gravity amplitude, which has already been confirmed in \citep{Carrasco2012}. Moreover, it is not hard to check that in $5$-pt case, the resulted expression can even be directly re-summed to be a formula purely consisted of MHV pentagons again
\begin{equation}\label{eq:newGRformula5pt}
  \mM^{(1)}_5=\sum_{\sump}\frac{Q(\s_1\s_2\s_3\s_4\s_5)}{\la\s_1\s_2\s_3\s_4\s_5\ra}\mP_5(\s_1\s_2\s_3\s_4\s_5).
\end{equation}

\spacesection

\section{From Yang-Mills to Gravity: $6$-pt $1$-Loop Amplitudes}
\label{sec:6ptcase}

Since in $5$-pt case we have observed that the MHV polygons serve as a very nice basis for $\mN=8$ supergravity amplitude, we would like to see whether they continue to work for more particles. But we need to start by checking $6$-pt case, since redundancies start to occur in the reduction formula of MHV hexagons. The strategy is still to expand the MHV polygons into loop integrals, and seek for color-kinematics duality, and once this duality is satisfied, we can directly check whether \prescription{} gives the correct gravity amplitude. However, one needs to be cautious, because due to the redundancies the expansion does not in general have the correct color structure (in the sense that it matches with the corresponding trivalent diagram, upon which color-kinematics duality is based \citep{Bern2008,Bern2010a,Carrasco2012}. This is always true at tree level, but one needs to take loop propagators into consideration at loop levels. For more detailed discussion, please refer to Appendix \ref{app:CKDuality}). In $6$-pt case it turns out that, once we tune the expansion to have the correct color structure under our construction, color-kinematics duality is just a subsequent outcome, although a new phenomena would arise at the same time, which we will discuss in later parts of the paper.

\subsection{Analysis of the Redundancies}

In Section \ref{sec:polygon}, we have already obtained a reduction formula \eqref{eq:fullhexreduction} for MHV hexagons, which is manifestly cyclic symmetric. However, \eqref{eq:fullhexreduction} is still not nice enough to work with. Instead, we choose a particular point for the $\alpha$ parameter, and add deviations upon it
\begin{equation}\label{eq:parameterdeviation}
  \begin{split}
    \alpha(123456)&=-\frac{[12]\la23\ra[34]\la45\ra[56]\la61\ra}{\ep(123456)}+\Delta(123456),\\
    \alpha(234561)&=-\frac{[23]\la34\ra[45]\la56\ra[61]\la12\ra}{\ep(234561)}-\Delta(123456).
  \end{split}
\end{equation}
We can see the original constraint \eqref{eq:alphaconstraint} on the $\alpha$ parameter is automatically satisfied. The virtue of analyzing around this particular point will be clear in the next subsection. But still $\Delta(234561)$ has no definition, and since $\alpha(234561)$ and $\alpha(123456)$ are assumed to be related by permutations, we would expect $\alpha(234561)$ can be naturally associated with a parameter $\Delta(234561)$ in the same way. And so it is necessary to further impose the condition
\begin{equation}\label{eq:Deltasymmetry}
  \Delta(234561)=-\Delta(123456).
\end{equation}
Then $\Delta(abcdef)$ shares the same symmetry as $\ep(abcdef)$ under cyclic permutations.

With these adjustments, the reduction formula of the MHV hexagons still only have pentagon integrals and box integrals, but no hexagon integral~\footnote{We don't consider here adding any tensor structures to the loop integrals, {because by applying Carrasco and Johansson's ansatz \cite{Carrasco2012} with the most general tensor structures to $6$-pt case, one can check that once the full color-kinematics duality is assumed to hold, any tensors of the type $l^2l^2$ just vanish, and any tensors of the type $l^2$ are purely gauge redundancies, which can be set to zero. We will not discuss this any further since it is not relevant for our current purpose.}}. The hexagon integral should be introduced via the identities between hexagon integral and pentagon integrals, which can also be worked out systematically \citep{VanNeerven1984} (see also \citep{Binoth1999,Binoth2005}). Since in the amplitude, each MHV hexagon is dressed by the color factor corresponding to the maximal color loop with the same sequence of labels, in a particular MHV hexagon we would add identity that involves only one hexagon integral whose labels fall into the correct sequence, otherwise in the resulted expansion formula the structure of the kinematic factor and the color factor would be drastically different, which is what we want to get rid of.
Then take $\mP(123456)$ as an example, the identity that meets these requirements is unique
\begin{equation}\label{eq:hexagonpentagonidentity}
  \begin{split}
    I_6(1|2|3|4|5|6)-\sum_{\mathbb{Z}_6}\frac{\ep(3456)}{\ep(123456)}I_5(12|3|4|5|6)=0.
  \end{split}
\end{equation}
We will dress it with a coefficient $\gamma(123456)$. And by the consideration of the symmetry of the identity \eqref{eq:hexagonpentagonidentity}, this functional coefficient should satisfy
\begin{align}
  \label{eq:gammasymmetry1}\gamma(123456)&=\gamma(234561),\\
  \label{eq:gammasymmetry2}\gamma(123456)&=\gamma(654321).
\end{align}

\subsection{Condition for Color-Kinematics Duality}

In order not to deviate into too much technical details, we choose to summarize here only the main results in the analysis of color-kinematics duality, and put the remaining details in Appendix \ref{app:ColorStructure}, \ref{app:ConstrainFreedom} and \ref{app:loopDuality}.

As has been stated, in $6$-pt case it is no longer true that the resulted formula automatically has the correct color structure. Since color-kinematics duality is based on the correct color structure, we should first make sure this condition is satisfied. We start by temporarily setting the parameter $\Delta=0$ (so we only consider $\gamma$), and find that in the scalar loop integral expansion, box integrals of all types ($I_4(a|bc|d|ef)$, $I_4(ab|cd|e|f)$ and $I_4(abc|d|e|f)$) already have the correct color structure. Since every MHV hexagon gives rise to its corresponding unique hexagon integral, the color structure of each hexagon integrals is also already correct. Then by looking at the coefficient in front of each pentagon integral (e.g.~$I_5(12|3|4|5|6)$), we obtain a constraint relating $\gamma(123456)$ and $\gamma(213456)$
\begin{equation}\label{eq:gammaconstraint}
  \frac{\gamma(123456)}{\la123456\ra\ep(123456)}+\frac{\gamma(213456)}{\la213456\ra\ep(213456)}+\frac{[12][34][56]s_{12}s_{34}s_{45}s_{56}(3|4+5|6)^2}{\la12\ra\la34\ra\la56\ra\ep(123456)\ep(213456)\ep(3456)}=0.
\end{equation}
As has been mentioned in the introduction, by permuting the labels, we can get constraints on other pairs of $\gamma$'s related by transposition of two adjacent labels. So we can also regard \eqref{eq:gammaconstraint} as a single functional constraint equation that depends on the choice of label ordering.

Now we turn on the redundancies parameterized by $\Delta$, and upon the previous analysis we only need to look at the effects of additional terms. Then the box integrals immediately constrain the $\Delta$ parameter to have the following form
\begin{equation}\label{eq:kappadefinition}
  \Delta(abcdef)=\frac{\la abcdef\ra}{\ep(abcdef)}\kappa(abcdef),
\end{equation}
where $\kappa(abcdef)$ is completely symmetric under permutations of the labels, and so later on we will abbreviate it as $\kappa$. Under this condition, the hexagon integrals are still untouched, while the constraint \eqref{eq:gammaconstraint} from pentagon integrals now becomes
\begin{equation}\label{eq:gammaconstraintnew}
  \begin{split}
    \frac{\gamma(123456)}{\la123456\ra\ep(123456)}+\frac{\gamma(213456)}{\la213456\ra\ep(213456)}&\\
    +\frac{[12][34][56]s_{12}s_{34}s_{45}s_{56}(3|4+5|6)^2}{\la12\ra\la34\ra\la56\ra\ep(123456)\ep(213456)\ep(3456)}-\kappa\frac{s_{12}s_{34}s_{45}s_{56}(3|4+5|6)^2}{\ep(123456)\ep(213456)\ep(3456)}&=0.
  \end{split}
\end{equation}

Since under \eqref{eq:gammaconstraintnew} the correct color structure is guaranteed, we can go on to check the condition for color-kinematics duality. Interestingly, it turns out that the kinematic dual identities between the kinematic numerators of the hexagon integrals and pentagon integrals are exactly equivalent to \eqref{eq:gammaconstraint} or \eqref{eq:gammaconstraintnew} (depending on whether we set $\kappa=0$), and the dual identities between kinematic numerators of the pentagon integrals and box integrals also automatically hold as a result of (repeatedly) applying \eqref{eq:gammaconstraint} or \eqref{eq:gammaconstraintnew}. And the occurrence of $\kappa$ doesn't break the duality at all. In other words, starting from the Yang-Mills formula \eqref{eq:newYMformula} in $6$-pt case and introduce redundant parameters $\gamma$ and $\kappa$ in the way as described above, just by requiring that the formula should have the correct color structure in its expansion onto scalar loop integrals, we know that these parameters only need to satisfy a single functional constraint \eqref{eq:gammaconstraintnew} in addition to their own symmetries, and then color-kinematics duality just comes for free, almost the same as what happens at $5$ points!

\subsection{A Global Inconsistency}
Before going on, let us perform a consistency check on the constraint \eqref{eq:gammaconstraint} or \eqref{eq:gammaconstraintnew}. We can see that, for every constraint with a specific configuration of the particle labels, it relates exactly two $\gamma$'s which differ by exchanging only one pair of neighboring labels. This can also be interpreted as, whenever we exchange two neighboring labels in $\gamma$, it would acquire some additional contribution as given by the inhomogeneous term in this constraint. Since this operation of exchanging labels allow us to exhaust the entire label configuration space starting from any specific point in it, we may expect that if we start from a certain point and move step by step, and if in the end we move back to the same starting point (label configuration), all the additional contributions acquired during the middle should add up to zero. This is an important consistency check in guaranteeing that the entire set of $\gamma$ parameters (with all in-equivalent label configurations) have a solution in terms of ordinary kinematics data.

\begin{figure}[htbp]
	\centering
		\includegraphics{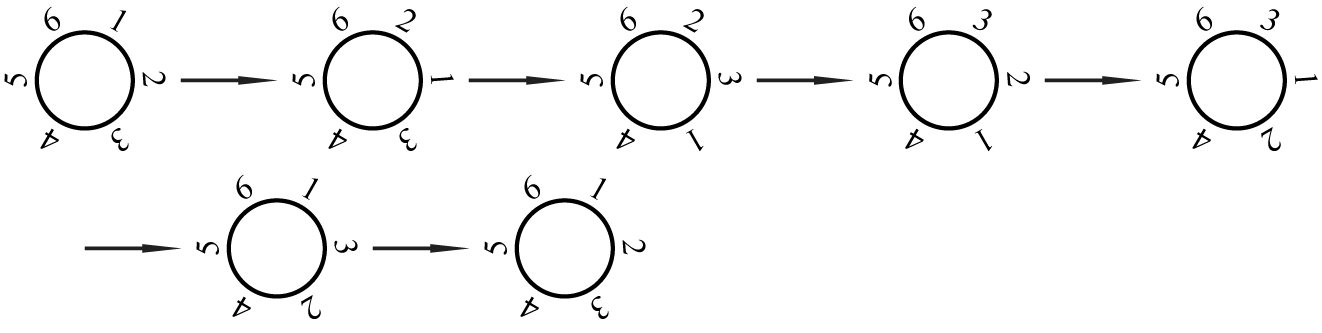}
	\caption{Consistent Moves}
	\label{fig:figure1}
\end{figure}

However, the constraint \eqref{eq:gammaconstraint} or \eqref{eq:gammaconstraintnew} that we have obtained in the previous subsection seems to be self-contradictory under this check. In more detail, the procedure as described above should be divided into two different classes. We may imagine the label arguments of $\gamma$ to reside on a circle, since by definition $\gamma$ should be invariant under cyclic permutation of the labels, and so there is no particular origin with respect to this operation. Then within the first class, no matter what kind of operation we do during the middle, after we have returned to the starting point, if the net effect is equivalent to that no labels have been brought to move around the circle, we would find that all the additional contributions ultimately cancel out and the $\gamma$ just returns to itself, which is consistent. An example is illustrated in Figure \ref{fig:figure1}.

\begin{figure}[htbp]
	\centering
		\includegraphics{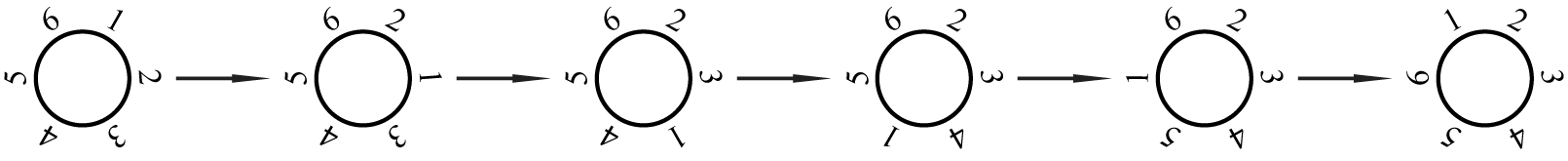}
	\caption{Inconsistent Moves}
	\label{fig:figure2}
\end{figure}

However, in the other class, as long as the net effect is equivalent to that at least one label has been brought around the circle (i.e.~let it to acquire a non-zero winding number), we would find a net discontinuity between the original $\gamma$ and the final $\gamma$
\begin{equation}\label{eq:gammaconsistency}
  \gamma(abcdef)_{\text{final}}=\gamma(abcdef)_{\text{original}}+(\text{additional term}).
\end{equation}
The simplest example of this kind is shown in Figure \ref{fig:figure2}.

Attention should be paid that, no matter whether the additional term in \eqref{eq:gammaconsistency} arise or not, its value depends only on the net effect (the winding numbers the labels acquired during the transformations), but not on the specific procedure of intermediate steps, and whenever a non-zero discontinuity is there, there is no means to tune $\kappa$ in order to eliminate it.

This indicates that, if we regard the $\gamma$ and $\kappa$ as being constructed with ordinary variables (such as with the usual spin-helicity formalism) and attempt to solve the entire set of constraints, we would end up in having no solution for them. However, this inconsistency in $\gamma$ is not so harmful as it appears to be. The reason is that, back in the beginning when $\gamma$ is first introduced, it comes with a purely mathematical identity which is completely transparent to physical analysis, and characterizes an auxiliary parameter space; especially, this parameter contributes nothing to any leading singularities. If we do not insist that this parameter be constructed by ordinary variables, the inconsistency is not really there. Although not globally solvable, these ``virtual'' parameters do encode physical data in an astonishingly neat and compact way via the functional constraint on them, and allow us to travel between scattering amplitudes in different theories. Or put it in another way, the auxiliary space here only serve as a context where the constraint (the only object which carries physical information) can be properly described. These would be clarified in detail in the upcoming analysis.

\spacesection

\section{New Formula for $6$-pt $1$-Loop MHV Amplitude in $\mN=8$ Supergravity}
\label{sec:6ptsugra}

From the previous section we know that although the existence of $\gamma$ and $\kappa$ in terms of ordinary kinematics data is at risk, the color-kinematics duality is very well satisfied, which still strongly tempts to suggest a ``gravity'' counterpart. Here in this section we are going to show that indeed we can still follow the original \prescription{} to obtain the correct formula for the corresponding $6$-pt $1$-loop gravity amplitude. This again confirms that these auxiliary parameters are not ``locally'' inconsistent. Instead, we should take the view that the constraints on these parameters work locally instead of globally, because all the physical information to be extracted only depends on certain subset of the constraints (or the same functional constraint with different arguments) which are mutually consistent with each other.

The same as in $5$-pt case, in $6$-pt case the formula obtained by \prescription{} can still be reformulated purely in terms of MHV hexagons, which is equivalent to the corresponding Yang-Mills amplitudes with the color factor in front of each MHV hexagon substituted by the kinematic coefficient of the hexagon scalar loop integral with the same ordering of labels in the loop integral expansion
\begin{equation}\label{eq:gravityamplitude6pt}
  \widetilde{\mathcal{M}}^{(1)}_6=\sum_{\sumh}\frac{\gamma(\s_1\s_2\ldots\s_6)}{\la\s_1\s_2\ldots\s_6\ra^2}\mP_6(\s_1\s_2\ldots\s_6).
\end{equation}

In order to check the consistency of \eqref{eq:gravityamplitude6pt} with the correct quadruple cuts in gravity, we only need to check two types (since we are dealing with MHV polygons, leading singularities in quadruple cuts of the type $(a|b|cd|ef)$ are trivially zero). Again we first assume $\kappa=0$. The first type is $(a|bc|d|ef)$. Without loss of generality, we can assign particular configuration of the labels. So in this case we study the cut $(1|23|4|56)$, which involves the following contributions
\begin{equation}
  \frac{\gamma(123456)}{\la123456\ra^2}+\frac{\gamma(123465)}{\la123465\ra^2}+\frac{\gamma(132456)}{\la132456\ra^2}+\frac{\gamma(132465)}{\la132465\ra^2}.
\end{equation}
The amazing fact is, we are still able to use the constraints \eqref{eq:gammasymmetry1}, \eqref{eq:gammasymmetry2} and \eqref{eq:gammaconstraint} repeatedly to fully eliminate the appearance of $\gamma$, and the final result is
\begin{equation}
  \begin{split}
    &\frac{[23]^2[45][61]s_{23}s_{45}s_{56}s_{61}(4|5+6|1)^2}{\la23\ra^2\la45\ra^2\la61\ra^2\la46\ra\la51\ra\ep(234561)\ep(324561)}\\
    -&\frac{[56][12][34]s_{56}s_{12}s_{23}s_{34}(1|2+3|4)^2}{\la123465\ra\la56\ra\la12\ra\la34\ra\ep(561234)\ep(1234)}-\frac{[56][13][24]s_{56}s_{13}s_{32}s_{24}(1|3+2|4)^2}{\la132465\ra\la56\ra\la13\ra\la24\ra\ep(561324)\ep(1324)}.
  \end{split}
\end{equation}
And this can actually be verified to be identical to the correct leading singularity of the gravity amplitude.

For the other type we choose to study $(1|2|3|456)$, which receives contributions from $6$ terms
\begin{equation}
  \frac{\gamma(123645)}{\la123645\ra^2}+\frac{\gamma(123654)}{\la123654\ra^2}+\frac{\gamma(123465)}{\la123465\ra^2}+\frac{\gamma(123564)}{\la123564\ra^2}+\frac{\gamma(123456)}{\la123456\ra^2}+\frac{\gamma(123546)}{\la123546\ra^2}.
\end{equation}
For this expression, we can use the same constraints in different ways to obtain different but equivalent final expressions, and still completely independent of $\gamma$, which can be shown to be equivalent. One example is
\begin{equation}
  \begin{split}
    &\left(\frac{\ep(123456)}{\la123456\ra}-\frac{[12]\ep(3456)}{\la12\ra\la34\ra\la65\ra\la36\ra\la45\ra}\right)\frac{[45][61][23]s_{45}s_{61}s_{12}s_{23}(6|1+2|3)^2}{\la45\ra\la61\ra\la23\ra\ep(123456)\ep(123546)\ep(6123)}\\
    -&\frac{[12]\ep(3456)}{\la12\ra\la34\ra\la65\ra\la36\ra\la45\ra}\frac{[56][12][34]s_{56}s_{12}s_{23}s_{34}(1|2+3|4)^2}{\la56\ra\la12\ra\la34\ra\ep(561234)\ep(651234)\ep(1234)}\\
    -&\frac{[12]\ep(3546)}{\la12\ra\la35\ra\la64\ra\la36\ra\la54\ra}\frac{[46][12][35]s_{46}s_{12}s_{23}s_{35}(1|2+3|5)^2}{\la46\ra\la12\ra\la35\ra\ep(461235)\ep(641235)\ep(1235)}\\
    +&\frac{[46][51][23]s_{46}s_{51}s_{12}s_{23}(5|1+2|3)^2}{\la123645\ra\la46\ra\la51\ra\la23\ra\ep(123465)\ep(5123)}+\frac{[56][41][23]s_{56}s_{41}s_{12}s_{23}(4|1+2|3)^2}{\la123654\ra\la56\ra\la41\ra\la23\ra\ep(123564)\ep(4123)},
  \end{split}
\end{equation}
which again has been checked numerically to be equivalent to the correct leading singularity.

Now we would also like to know how the $\kappa$ parameter would affect the leading singularities in the potential gravity amplitudes. The answer is that it has exactly zero effect. More explicitly, the quadruple cut $(1|23|4|56)$) receives an additional contribution
\begin{equation}
  \begin{split}
    \bigg\{\frac{s_{56}s_{12}s_{23}s_{34}(1|2+3|4)^2}{\la123465\ra\ep(561234)\ep(1234)}+\frac{s_{56}s_{13}s_{32}s_{24}(1|3+2|4)^2}{\la132465\ra\ep(561324)\ep(1324)}&\\
    -\frac{[23]\ep(4561)}{\la23\ra\la45\ra\la61\ra\la46\ra\la51\ra}\frac{s_{23}s_{45}s_{56}s_{61}(4|5+6|1)^2}{\ep(234561)\ep(324561)\ep(4561)}&\bigg\}\kappa,
  \end{split}
\end{equation}
which can be checked to be exactly zero. And the cut $(1|2|3|456)$ receives an additional contribution
\begin{equation}
  \begin{split}
    \bigg\{-\left(\frac{\ep(123456)}{\la123456\ra}-\frac{[12]\ep(3456)}{\la12\ra\la34\ra\la65\ra\la36\ra\la45\ra}\right)\frac{s_{45}s_{61}s_{12}s_{23}\ep(6123)}{\ep(123456)\ep(123546)\ep(6123)}&\\
    +\frac{[12]\ep(3456)}{\la12\ra\la34\ra\la65\ra\la36\ra\la45\ra}\frac{s_{56}s_{12}s_{23}s_{34}(1|2+3|4)^2}{\ep(561234)\ep(651234)\ep(1234)}&\\
    +\frac{[12]\ep(3546)}{\la12\ra\la35\ra\la64\ra\la36\ra\la54\ra}\frac{s_{46}s_{12}s_{23}s_{35}(1|2+3|5)^2}{\ep(461235)\ep(641235)\ep(1235)}&\\
    -\frac{s_{46}s_{51}s_{12}s_{23}(5|1+2|3)^2}{\la123645\ra\la123465\ra\ep(5123)}-\frac{s_{56}s_{41}s_{12}s_{23}(4|1+2|3)^2}{\la123654\ra\ep(123564)\ep(4123)}&\bigg\}\kappa,
  \end{split}
\end{equation}
which also vanishes. As a result, in both types of quadruple cuts, the terms that contain $\kappa$ just cancel out completely. Hence we do not only get a formula \eqref{eq:gravityamplitude6pt} with the correct leading singularities of gravity amplitudes, but actually a one parameter family of them. The price for this is that we have to build the color-kinematics duality on a virtual auxiliary space.

\spacesection

\section{Discussions}
\label{conclusion}

As a brief summary, we used MHV polygons to build $1$-loop amplitudes for any number of particles in $\mN=4$ super Yang-Mills theory, and up to $6$ points in $\mN=8$ supergravity, all of which are expressed as a simple sum over terms enjoying manifest cyclic and reflection invariance and are related purely by permutations. In other words, with this basis, the amplitudes can be generated with a single functional object (either a functional coefficient or a functional constraint).

We have observed that non-trivial structures start to appear at $6$ points. The coefficients multiplying the MHV hexagons cannot be expressed in terms of ordinary kinematics data. Instead, the amplitude here is formulated by imposing upon these coefficients a set of linear constraints, all of which have the same form (or, there is just a single functional constraint), which totally determine all the physical information needed from the amplitude. The striking fact is that, since these redundant parameters are not globally solvable, we haven't really obtained the real color-kinematics duality on the Yang-Mills side in our specific construction. However, the result shows that we can still circumvent it to get the correct gravity amplitude, just by imposing the algebraic relations but not solving them. By origin, $\gamma$ and $\kappa$ are only parameters that describe an auxiliary space. And so they can be regarded as virtual, and their only purpose is to form a context to formulate the constraint. It is really the constraint among these auxiliary variables that generates all needed physical data.

As a natural generalization to the work presented in this paper, we can directly boost the form in \eqref{eq:gravityamplitude6pt} to a conjecture for any number of particles
\begin{equation}
  \widetilde{\mathcal{M}}^{(1)}_n=\sum_{\sumn}\frac{\gamma(\s_1\s_2\ldots\s_n)}{\la\s_1\s_2\ldots\s_n\ra^2}\mP_n(\s_1\s_2\ldots\s_n),
\end{equation}
with the $\gamma(\s_1\s_2\ldots\s_n)$ still fully cyclic and reflection invariant but in general not solvable. Instead, we need to find out the constraints forced onto these coefficients, which are mutually consistent with respect to each quadruple cut but not necessarily consistent as a whole. By analogy, the coefficients here are still purely redundant parameters coming from adding kinematic identities into the expansion of MHV polygons in analyzing the color-kinematics duality (here by ``purely redundant'' we mean the parameters are dressed with pure zeros). We have seen that given the color-kinematics duality is satisfied, the gravity amplitude can be obtained just simply by substituting the color factor in front of each MHV polygons with the coefficient of the highest order scalar loop integrals in the expansion of the polygons. But in order to achieve that, at this point we still have to go back to the complicated expansions in terms of the scalar loop integrals and carefully fine-tune the redundancies. Especially, we have to seek for a manifestly symmetric decomposition of the MHV polygons into scalar loop integrals by introducing pure redundancies in a proper way, which in general seems hard. But most of the information in this analysis is actually not needed in the final answer. What we expect instead is a method that can work out the single coefficient needed to put in front of the MHV polygons or the constraints they should satisfy, without really working with the full set of loop integrals. Moreover, although the entire set of constraints are inconsistent, they have some good behaviors under the operations as described before, which seems to suggest there is some non-trivial mathematical structure to be better understood for this auxiliary space. And we hope that once this auxiliary space is described in a better way, it may help reach a general formula for $1$-loop MHV amplitudes in $\mN=8$ supergravity.

\spacesection

\appendix

\section{A Brief Review of Color-Kinematics Duality}\label{app:CKDuality}

In brief, color-kinematics duality refers to the situation that in a formula where the Yang-Mills amplitude is expressed totally in terms of trivalent diagrams (whose structures have one-to-one correspondence to the color structures in terms of $f_{abc}$), for every color Jacobi identity
\begin{equation}
  c_i\pm c_j\pm c_k=0,
\end{equation}
there exists a dual kinematic identity
\begin{equation}
  n_i\pm n_j\pm n_k=0
\end{equation}
sharing the same algebraic structure, but with each color factor substituted by its corresponding kinematic numerator \citep{Bern2008} (Notice that this is not necessarily a rational function. The remaining kinematic factor is a product of propagators indicated by the trivalent diagram).

At tree level, when this duality is satisfied, the amplitude can in general be written in the form
\begin{equation}\label{a:YMCKDualityTree}
  \mA^{(0)}_n=\sum_i\frac{c_in_i}{\left(\prod{s}\right)_i},
\end{equation}
where the summation is over all different trivalent diagrams. As the simplest example, the $4$-pt tree-level Yang-Mills amplitude can be written as (without the helicity factor)
\begin{equation}
  \tmA^{(0)}_4=\frac{f_{12a}f_{a34}}{s_{12}}\frac{[12]}{\la23\ra\la34\ra\la41\ra}+\frac{f_{13a}f_{a24}}{s_{13}}\frac{[13]}{\la32\ra\la24\ra\la41\ra},
\end{equation}
which already satisfies the color-kinematics duality, with the dual kinematic identity associated to its unique color Jacobi identity
\begin{equation}
  \frac{[12]}{\la23\ra\la34\ra\la41\ra}-\frac{[13]}{\la32\ra\la24\ra\la41\ra}-0=0.
\end{equation}
$4$-pt is special in that, the duality holds for whatever form the amplitude is written in, which is due to the possibility of adding a redundancy parameterized by $\alpha$
\begin{equation}
  \tmA^{(0)}=f_{12a}f_{a34}\left(\frac{1}{\la12\ra\la23\ra\la34\ra\la41\ra}+\alpha\right)+f_{13a}f_{a24}\left(\frac{1}{\la13\ra\la32\ra\la24\ra\la41\ra}-\alpha\right)-f_{1a4}f_{a23}\alpha
\end{equation}
while preserving the validity of the dual kinematic identity. For higher number of particles the formula for any amplitude does not automatically satisfy this duality, but the formula that meets this requirement is in general not unique.

When a formula for tree-level Yang-Mills amplitude satisfies color-kinematics duality, Bern, Carrasco and Johansson conjectured that the correct formula for gravity with the same number of particles can be directly obtained by substituting in \eqref{a:YMCKDualityTree} each color factor $c_i$ by another copy of its corresponding kinematic numerator $n_i$ \citep{Bern2008}
\begin{equation}
  \mM^{(0)}_n=\sum_i\frac{\widetilde{n}_in_i}{\left(\prod{s}\right)_i}.
\end{equation}
This is called double-copy construction. Especially, the two copies of kinematic numerator can come from two different formulas that satisfy color-kinematics duality respectively. This conjecture has been proved at tree level \citep{Bern2010}.

The story is largely the same at loop level, except that here the duality is claimed to hold down to the integrand \citep{Bern2010a}. In other words, the amplitude can be expressed as
\begin{equation}\label{a:YMCKDualityLoop1}
  \mA^{(L)}_n=\sum_i\int\left(\prod{d^4k}\right)\frac{c_in_i(\{p\},\{k\})}{\left(\prod{s}\right)_i\left(\prod{l^2}\right)_i},
\end{equation}
where the summation is still over all different trivalent diagrams, $\{k\}$ is the set of loop momenta to be integrated over, $\prod{l^2}$ is the product of loop propagators, and $\prod{s}$ is the product of propagators corresponding to additional tree structures of the trivalent diagram (if there is any). Note that, at loop level the kinematic numerator $n_i(\{p\},\{k\})$ is in general a function of both the external on-shell momenta and the loop momenta, so it may acquire tensor structures (in terms of loop momenta). The way to go to gravity is still to substitute each color factor $c_i$ by another copy of the kinematic numerator $\widetilde{n}_i(\{p\},\{k\})$
\begin{equation}
  \mM^{(L)}_n=\sum_i\int\left(\prod{d^4k}\right)\frac{\widetilde{n}_i(\{p\},\{k\})n_i(\{p\},\{k\})}{\left(\prod{s}\right)_i\left(\prod{l^2}\right)_i},
\end{equation}

However, if it is known that any tensor structures appearing in $n_i$ can be gauged away (such as in $5$-pt case as discussed by Carrasco and Johansson), then we can bring the color factor together with its corresponding kinematic numerator out of the integration in \eqref{a:YMCKDualityLoop1}
\begin{equation}\label{a:YMCKDualityLoop2}
  \mA^{(L)}_n=\sum_ic_in_i(\{p\})\int\left(\prod{d^4k}\right)\frac{1}{\left(\prod{s}\right)_i\left(\prod{l^2}\right)_i}=\sum_i\frac{c_in_i}{\left(\prod{s}\right)_i}I_i,
\end{equation}
where $I$ denotes the ordinary scalar loop integral corresponding to the trivalent diagrams (without the propagators from the tree structures). And the conjectured gravity amplitude is
\begin{equation}\label{a:GRCKDualityLoop2}
  \mM^{(L)}_n=\sum_i\frac{\widetilde{n}_in_i}{\left(\prod{s}\right)_i}I_i.
\end{equation}
Since in the cases that we study here the tensor structures are still redundant, we always use \eqref{a:YMCKDualityLoop2} and \eqref{a:GRCKDualityLoop2} when analyzing color-kinematics duality.

\spacesection

\section{Analysis of the Color Structure}\label{app:ColorStructure}

This appendix summarizes the detailed analysis on the condition for $6$-pt Yang-Mills amplitude to have the correct color structure when MHV hexagons are expanded into scalar loop integrals. To simplify analysis, we temporarily set the parameter $\Delta=0$. Then in the hexagon reduction formula, box integrals of the type $I_4(a|bc|d|ef)$ are
\begin{equation}\label{a:boxintegral1}
  -\sum_{\mathbb{Z}_6}\frac{\la12\ra[23]\la34\ra[41]\la45\ra[56]\la61\ra[14](1|2+3|4)^2}{\ep(1234)\ep(4561)}I_4(1|23|4|56).
\end{equation}
In order to obtain the correct color factor for this type of box integral, since this color factor is a linear combination of four different color factors of the highest level, we should expect that in the  expansion of the full amplitude (multiply the above by the corresponding Parke-Taylor form and then sum over MHV hexagons), the kinematic factors in front of $I_4(a|bc|d|ef)$, $I_4(a|bc|d|fe)$, $I_4(a|cb|d|ef)$ and $I_4(a|cd|d|fe)$ should be equivalent up to a possible minus sign, and this minus sign occurs when the two integrals differ by exchanging labels only within either $(bc)$ or $(ef)$. It is straightforward to check that \eqref{a:boxintegral1} meets this requirement, so this type of box integrals already have the correct color structure at the particular point as chosen in \eqref{eq:parameterdeviation} when $\Delta=0$. And then it is also easy to see the same goes with the box integrals of the type $I_4(ab|cd|e|f)$, which are
\begin{equation}\label{a:boxintegral2}
  -\sum_{\mathbb{Z}_6}\frac{[12]\la23\ra[34]\la45\ra[56]\la61\ra s_{56}(5|3+4|6)^2}{\ep(3456)\ep(5612)}I_4(12|34|5|6).
\end{equation}

Then for the third type $I_4(abc|d|e|f)$, the combined formula is
\begin{equation}\label{a:boxintegral3}
  \begin{split}
    \sum_{\mathbb{Z}_6}\frac{s_{45}s_{56}\la34\ra[46]\la61\ra}{\ep(123456)}\bigg(&-\frac{[12]\la23\ra[34]\la45\ra[56][3|4+5|6\ra}{\ep(3456)}\\
    &+\frac{\la12\ra[23][45]\la56\ra[61]\la4|5+6|1]}{\ep(4561)}\bigg)I_4(123|4|5|6).
  \end{split}
\end{equation}
The analysis is a little bit different for this case because of the color structure. For example, this particular box integral $I_4(123|4|5|6)$ would give contributions to two color structures $f_{a1b}f_{b23}$ and $f_{a3b}f_{b12}$. So for each coefficient of this type of box integrals, we need to separate it into two parts and compare each part with the corresponding one from other box integrals that would contribute to the same box color structure, which is hard to be carried out. However, a careful look at the decomposition of color factors shows that another box integral $I_4(321|4|5|6)$ exactly and only gives contributions to the same two color structures, and with the same sign. By the fact that for each color structure, the decomposed coefficient should match between the two box integrals, the two entire coefficients should also be equivalent, and this has been verified by numerical check. Although this hasn't fully shown that the correct color structure can be obtained for this type of box integral, it is already a strong consistency check. And later in the analysis of color-kinematics duality, we'll further strengthen this argument.

Then what is remaining are the pentagon integrals, which will give non-trivial constraints. The hexagon and pentagon integrals in the reduction formula are
\begin{equation}\label{a:hexagonpentagonintegral}
  \begin{split}
    \gamma(123456)&I_6(1|2|3|4|5|6)\\
    +\sum_{\mathbb{Z}_6}\left\{\frac{[12]\la23\ra[34]\la45\ra[56]\la61\ra s_{34}s_{45}s_{56}(3|4+5|6)^2}{\ep(123456)\ep(3456)}-\gamma(123456)\frac{\ep(3456)}{\ep(123456)}\right\}&I_5(12|3|4|5|6).
  \end{split}
\end{equation}
In order that the pentagon integrals have the correct color structure, we should have
\begin{equation}
  \begin{split}
    0&=\left\{\frac{[12]\la23\ra[34]\la45\ra[56]\la61\ra s_{34}s_{45}s_{56}(3|4+5|6)^2}{\ep(123456)\ep(3456)}-\gamma(123456)\frac{\ep(3456)}{\ep(123456)}\right\}\frac{1}{\la123456\ra}\\
    &+\left\{\frac{[21]\la13\ra[34]\la45\ra[56]\la62\ra s_{34}s_{45}s_{56}(3|4+5|6)^2}{\ep(213456)\ep(3456)}-\gamma(213456)\frac{\ep(3456)}{\ep(213456)}\right\}\frac{1}{\la213456\ra},
  \end{split}
\end{equation}
which gives
\begin{equation}\label{a:gammaconstraint}
  \frac{[12][34][56]s_{12}s_{34}s_{45}s_{56}(3|4+5|6)^2}{\la12\ra\la34\ra\la56\ra\ep(123456)\ep(213456)\ep(3456)}+\frac{\gamma(123456)}{\la123456\ra\ep(123456)}+\frac{\gamma(213456)}{\la213456\ra\ep(213456)}=0.
\end{equation}
The constraints on $\gamma$ with the other sequences of labels are related just by permutations. So we can regard \eqref{a:gammaconstraint} either as a single functional constraint equation, or together with its permutations forming a set of constraints on the set of $\gamma$'s with all possible orderings of particle labels.

\spacesection

\section{Effects of the Parameter $\Delta$}\label{app:ConstrainFreedom}
This appendix complements Appendix \ref{app:ColorStructure} in studying the effect of $\Delta$ on the color structures. When $\Delta$ is non-zero, the terms corresponding to the type $I_4(a|bc|d|ef)$ receive additional contributions as
\begin{equation}\label{a:boxintegral1a}
  -\sum_{\mathbb{Z}_6}\frac{s_{14}\ep(123456)(1|2+3|4)^2}{\ep(1234)\ep(4561)}\Delta(123456)I_4(1|23|4|56).
\end{equation}
Since the original part itself already gives the desired color factor, if this new expression should preserve this property, the additional term containing $\Delta$ should satisfy
\begin{equation}
  -\frac{s_{14}\ep(123456)(1|2+3|4)^2}{\ep(1234)\ep(4561)}\frac{\Delta(123456)}{\la123456\ra}-\frac{s_{14}\ep(132456)(1|3+2|4)^2}{\ep(1324)\ep(4561)}\frac{\Delta(132456)}{\la132456\ra}=0,
\end{equation}
which is simplified to
\begin{equation}\label{a:Deltaconstraint1}
  \frac{\ep(123456)\Delta(123456)}{\la123456\ra}=\frac{\ep(132456)\Delta(132456)}{\la132456\ra}.
\end{equation}
This already gives a very strong constraint on $\Delta$. But to further confirm it, take another type of box integral $I_4(ab|cd|e|f)$
\begin{equation}\label{a:boxintegral2a}
  \begin{split}
    \sum_{\mathbb{Z}_6}\frac{s_{56}\ep(123456)(5|3+4|6)^2}{\ep(3456)\ep(5612)}\Delta(123456)I_4(12|34|5|6).
  \end{split}
\end{equation}
Again by requiring the desired color factor for this box integral, we would get
\begin{equation}\label{a:Deltaconstraint2}
  \frac{\ep(123456)\Delta(123456)}{\la123456\ra}=\frac{\ep(213456)\Delta(213456)}{\la213456\ra}.
\end{equation}
We may observe that given the symmetry of $\Delta$, \eqref{a:Deltaconstraint2} is compatible with \eqref{a:Deltaconstraint1}, and they together indicate that the particular form
\begin{equation}\label{a:kappadefinition}
  \kappa=\frac{\ep(abcdef)\Delta(abcdef)}{\la abcdef\ra}
\end{equation}
should be invariant under all permutations of the labels. In terms of the new parameter $\kappa$, terms involving the type $I_4(abc|d|e|f)$ acquires the additional contribution
\begin{equation}\label{a:boxintegral3a}
  \kappa\sum_{\mathbb{Z}_6}\frac{s_{45}s_{56}[46]\la123456\ra}{\ep(123456)}\left(\frac{\la34\ra[3|4+5|6\ra}{\ep(3456)}-\frac{\la61\ra\la4|5+6|1]}{\ep(4561)}\right)I_4(123|4|5|6).
\end{equation}
Numerically, take $I_4(123|4|5|6)$ for example, its additional kinematic factor is invariant under the exchange of label $1$ and $3$, indicating that $\kappa$ doesn't break the correct color structure for this type of box integral as well.

The pentagon integrals in the reduction formula now takes the form
\begin{equation}\label{a:pentagonintegralkappa}
  \begin{split}
    \sum_{\mathbb{Z}_6}\bigg\{&\frac{[12]\la23\ra[34]\la45\ra[56]\la61\ra s_{34}s_{45}s_{56}(3|4+5|6)^2}{\ep(123456)\ep(3456)}\\
    -&\kappa\frac{\la123456\ra s_{34}s_{45}s_{56}(3|4+5|6)^2}{\ep(123456)\ep(3456)}-\gamma(123456)\frac{\ep(3456)}{\ep(123456)}\bigg\}I_5(12|3|4|5|6).
  \end{split}
\end{equation}
Due to this additional term involving $\kappa$, if we still want to preserve the correct color structure for pentagon integrals, the original constraint \eqref{a:gammaconstraint} on $\gamma$ should be modified to
\begin{equation}\label{a:gammaconstraintnew}
  \begin{split}
    \frac{[12][34][56]s_{12}s_{34}s_{45}s_{56}(3|4+5|6)^2}{\la12\ra\la34\ra\la56\ra\ep(123456)\ep(213456)\ep(3456)}-\kappa\frac{s_{12}s_{34}s_{45}s_{56}(3|4+5|6)^2}{\ep(123456)\ep(213456)\ep(3456)}\\
    +\frac{\gamma(123456)}{\la123456\ra\ep(123456)}+\frac{\gamma(213456)}{\la213456\ra\ep(213456)}&=0.
  \end{split}
\end{equation}

\spacesection

\section{Analysis of Color-Kinematics Duality}\label{app:loopDuality}

This appendix studies the condition for color-kinematics duality when the scalar loop integral expansion of $6$-pt Yang-Mills amplitude we have obtained has the correct color structure.

Again let us set $\kappa=0$, and we first look into the relations between hexagon integrals and pentagon integrals. In order that color-kinematics duality holds at this level, for example, from the pentagon integral $I_5(12|3|4|5|6)$, together with two hexagon integrals $I_6(1|2|3|4|5|6)$ and $I_6(2|1|3|4|5|6)$, we should have the following constraint
\begin{equation}\label{a:pentagonBCJ}
  \begin{split}
    &\left\{\frac{[12]\la23\ra[34]\la45\ra[56]\la61\ra s_{34}s_{45}s_{56}(3|4+5|6)^2}{\ep(123456)\ep(3456)}-\gamma(123456)\frac{\ep(3456)}{\ep(123456)}\right\}\frac{1}{\la123456\ra}\\
    &=\left\{\frac{\gamma(123456)}{\la123456\ra}-\frac{\gamma(213456)}{\la213456\ra}\right\}\frac{1}{s_{12}}.
  \end{split}
\end{equation}
Then by using the identity
\begin{equation}
  \ep(123456)+\ep(213456)+s_{12}\ep(3456)=0,
\end{equation}
it is easy to show that \eqref{a:pentagonBCJ} is actually equivalent to the constraint \eqref{a:gammaconstraint} on $\gamma$ from the analysis of color structure.

For the box integrals of the type $I_4(a|bc|d|ef)$, we pick up $I_4(1|23|4|56)$ as an example, and check the relation between this and $I_5(1|2|3|4|56)$ and $I_5(1|3|2|4|56)$. Substraction of the coefficients of the two pentagon integrals gives
\begin{equation}
  \begin{split}
    &\left\{\frac{[56]\la61\ra[12]\la23\ra[34]\la45\ra s_{12}s_{23}s_{34}(1|2+3|4)^2}{\ep(561234)\ep(1234)}-\frac{\ep(1234)\gamma(561234)}{\ep(561234)}\right\}\frac{1}{\la561234\ra}\\
    -&\left\{\frac{[56]\la61\ra[13]\la32\ra[24]\la45\ra s_{13}s_{23}s_{24}(1|2+3|4)^2}{\ep(561324)\ep(1324)}-\frac{\ep(1324)\gamma(561324)}{\ep(561324)}\right\}\frac{1}{\la561324\ra},
  \end{split}
\end{equation}
which, after imposing the constraint \eqref{a:gammaconstraint} on $\gamma$ from the pentagon color factors, results in an expression that is entirely independent of $\gamma$
\begin{equation}\label{a:boxintegralBCJ1}
  \begin{split}
    +&\frac{[56][12][34]s_{12}s_{23}s_{34}(1|2+3|4)^2}{\la56\ra\la12\ra\la34\ra\ep(561234)\ep(1234)}+\frac{[56][13][24]s_{13}s_{23}s_{24}(1|2+3|4)^2}{\la56\ra\la13\ra\la24\ra\ep(561324)\ep(1234)}\\
    -&\frac{[23][45][61]s_{23}s_{45}s_{56}s_{61}(4|5+6|1)^2\ep(1234)}{\la23\ra\la45\ra\la61\ra\ep(561234)\ep(561324)\ep(4561)}.
  \end{split}
\end{equation}
If color-kinematics duality is to hold between this type of box integrals and the pentagon integrals, we should expect the above expression to be equivalent to the coefficient of the box times $s_{23}$
\begin{equation}
  -\frac{\la12\ra[23]\la34\ra[41]\la45\ra[56]\la61\ra[14](1|2+3|4)^2}{\ep(1234)\ep(4561)\la123456\ra}\cdot s_{23}.
\end{equation}
This has been verified with numerical values. Given this is satisfied, the other dual kinematics relations in the same category are automatic just by permuting the labels. The interesting fact here is that we don't even need to use the explicit form of the coefficients of the hexagon integrals.

Then go on to study the type $I_4(ab|cd|e|f)$. For this we pick up $I_4(12|34|5|6)$ together with $I_5(1|2|34|5|6)$ and $I_5(2|1|34|5|6)$. Substraction of the coefficients of the two pentagon integrals gives
\begin{equation}
  \begin{split}
    &\left\{\frac{[34]\la45\ra[56]\la61\ra[12]\la23\ra s_{56}s_{61}s_{12}(5|6+1|2)^2}{\ep(345612)\ep(5612)}-\frac{\ep(5612)}{\ep(345612)}\gamma(345612)\right\}\frac{1}{\la345612\ra}\\
    -&\left\{\frac{[34]\la45\ra[56]\la62\ra[21]\la13\ra s_{56}s_{62}s_{12}(5|6+2|1)^2}{\ep(345621)\ep(5621)}-\frac{\ep(5621)}{\ep(345621)}\gamma(345621)\right\}\frac{1}{\la345621\ra}.
  \end{split}
\end{equation}
Similarly, after imposing the constraints \eqref{a:gammaconstraint}, this reduces to another $\gamma$-independent expression
\begin{equation}\label{a:boxintegralBCJ2}
  \begin{split}
    &\frac{[34][56][12]s_{56}s_{61}s_{12}(5|6+1|2)^2}{\la34\ra\la56\ra\la12\ra\ep(345612)\ep(5612)}+\frac{[34][56][12]s_{56}s_{62}s_{12}(5|6+2|1)^2}{\la34\ra\la56\ra\la12\ra\ep(345621)\ep(5612)}\\
    +&\frac{[12][34][56]s_{12}s_{34}s_{45}s_{56}(3|4+5|6)^2\ep(5612)}{\la12\ra\la34\ra\la56\ra\ep(345612)\ep(345621)\ep(3456)},
  \end{split}
\end{equation}
which has been numerically checked to be equivalent to the coefficient of $I_4(12|34|5|6)$ times $s_{12}$
\begin{equation}
  -\frac{[12]\la23\ra[34]\la45\ra[56]\la61\ra s_{56}(5|3+4|6)^2}{\ep(3456)\ep(5612)\la123456\ra}\cdot s_{12}.
\end{equation}
This confirms that with the condition of correct color structure, color-kinematics duality also holds between pentagon integrals and the box integral of this second type.

For the remaining type $I_4(abc|d|e|f)$, we choose to study $I_4(123|4|5|6)$, and we would like to check whether the kinematic factor of this box integral with color structure $f_{123456}$ is consistent with color-kinematics duality in relation to the pentagon integrals. As has be mentioned before, the color structure $f_{123456}$ contributes to two box color structures, so we should study the dual kinematics relations for both structures, and only the sum (or substraction depending on the definition of the color structure) of the two results will equal the kinematic factor of $I_4(123|4|5|6)$.

For the first color structure $f_{a1b}f_{b23}$, its related pentagon integrals are $I_5(1|23|4|5|6)$ and $I_5(23|1|4|5|6)$. The substraction of the two is
\begin{equation}
  \begin{split}
    &\left\{\frac{[23]\la34\ra[45]\la56\ra[61]\la12\ra s_{45}s_{56}s_{61}(4|5+6|1)^2}{\ep(234561)\ep(4561)}-\frac{\ep(4561)}{\ep(234561)}\gamma(234561)\right\}\frac{1}{\la234561\ra}\\
    -&\left\{\frac{[23]\la31\ra[14]\la45\ra[56]\la62\ra s_{14}s_{45}s_{56}(1|4+5|6)^2}{\ep(231456)\ep(1456)}-\frac{\ep(1456)}{\ep(231456)}\gamma(231456)\right\}\frac{1}{\la231456\ra}.
  \end{split}
\end{equation}
Here we are still able to fully eliminate the $\gamma$, but should use two constraints. In the end, we would obtain the following expression
\begin{equation}\label{a:boxintegralBCJ3part1}
  \begin{split}
    &\frac{[23][45][61]s_{45}s_{56}s_{61}(4|5+6|1)^2}{\la23\ra\la45\ra\la61\ra\ep(234561)\ep(4561)}+\frac{[23][14][56]s_{14}s_{45}s_{56}(1|4+5|6)^2}{\la23\ra\la14\ra\la56\ra\ep(231456)\ep(4561)}\\
    -&\frac{[12][34][56]s_{12}s_{34}s_{45}s_{56}(3|4+5|6)^2\ep(4561)}{\la12\ra\la34\ra\la56\ra\ep(123456)\ep(213456)\ep(3456)}-\frac{[13][45][62]s_{13}s_{45}s_{56}s_{62}(4|5+6|2)^2\ep(4561)}{\la13\ra\la45\ra\la62\ra\ep(134562)\ep(314562)\ep(4562)}.
  \end{split}
\end{equation}
For the other color structure $f_{a3b}f_{b12}$, we can obtain the formula just by doing the permutation $\{1\rightarrow3,2\rightarrow1,3\rightarrow2\}$ to the above expression, which is
\begin{equation}\label{a:boxintegralBCJ3part2}
  \begin{split}
    &\frac{[12][45][63]s_{45}s_{56}s_{63}(4|5+6|3)^2}{\la12\ra\la45\ra\la63\ra\ep(124563)\ep(4563)}+\frac{[12][34][56]s_{34}s_{45}s_{56}(3|4+5|6)^2}{\la12\ra\la34\ra\la56\ra\ep(123456)\ep(4563)}\\
    -&\frac{[31][24][56]s_{31}s_{24}s_{45}s_{56}(2|4+5|6)^2\ep(4563)}{\la31\ra\la24\ra\la56\ra\ep(312456)\ep(132456)\ep(2456)}-\frac{[32][45][61]s_{32}s_{45}s_{56}s_{61}(4|5+6|1)^2\ep(4563)}{\la32\ra\la45\ra\la61\ra\ep(324561)\ep(234561)\ep(4561)}.
  \end{split}
\end{equation}
Then, since the color structures under study are decomposed as
\begin{equation}\label{a:boxintegral3color}
  \begin{split}
    f_{abd}f_{b1c}f_{c23}&=f_{a1b}f_{b2c}f_{c3d}-f_{a1b}f_{b3c}f_{c2d}-f_{a2b}f_{b3c}f_{c1d}+f_{a3b}f_{b2c}f_{c1d},\\
    f_{abd}f_{b3c}f_{c12}&=f_{a3b}f_{b1c}f_{c2d}-f_{a3b}f_{b2c}f_{c1d}-f_{a1b}f_{b2c}f_{c3d}+f_{a2b}f_{b1c}f_{c3d},
  \end{split}
\end{equation}
it is \eqref{a:boxintegralBCJ3part1}$-$\eqref{a:boxintegralBCJ3part2} that should be equivalent to the kinematic factor of the box integral $I_4(123|4|5|6)$ multiplied by $s_{123}=(p_1+p_2+p_3)^2$
\begin{equation}
  \begin{split}
    \frac{s_{45}s_{56}[46]}{\ep(123456)}\bigg(&-\frac{[12]\la23\ra[34]\la45\ra[56]\la61\ra\la34\ra[3|4+5|6\ra}{\ep(3456)}\\
    &+\frac{\la12\ra[23]\la34\ra[45]\la56\ra[61]\la61\ra\la4|5+6|1]}{\ep(4561)}\bigg)\cdot s_{123}.
  \end{split}
\end{equation}
This equivalence has also been confirmed numerically. In doing this, we've not only verified the color-kinematics duality for this case, but also obtained the desired decomposition of the kinematic factor in \eqref{a:boxintegral3} into \eqref{a:boxintegralBCJ3part1} and \eqref{a:boxintegralBCJ3part2} for $I_4(abc|d|e|f)$ in order to
give rise to the desired color structure for this type of box integrals, thus filling in the gap left by previous analysis.

So far we see that when $\kappa=0$, the color-kinematics duality is automatically satisfied. Now we turn on $\kappa$, and check whether this parameter would break the duality. Here we should be more careful, because non-vanishing $\kappa$ would modify the constraint on $\gamma$, and so in checking the duality, wherever $\gamma$ appears, we should count the additional term in \eqref{a:gammaconstraintnew}.

We still check the examples as we choose before, and only focus on the additional terms that would appear. At the level between hexagons and pentagons, if we add the $\kappa$ term from the coefficient of the pentagon into the L.H.S.~of the original relation \eqref{a:pentagonBCJ}, we can observe that the modified relation is again just the condition of correct color structure for pentagons with $\kappa$ turned on. So the duality at this level is un-touched.

At the level between pentagon integrals and box integrals, we first go back to the type involving $I_4(a|bc|d|ef)$ (e.g.~$I_4(1|23|4|56)$). Now the coefficient of the pentagon integrals $I_5(1|2|3|4|56)$ and $I_5(1|3|2|4|56)$ together with the constraint on $\gamma$ would add in \eqref{a:boxintegralBCJ1} the following additional terms
\begin{equation}\label{a:boxintegralBCJ1new}
  -\frac{s_{12}s_{23}s_{34}(1|2+3|4)^2}{\ep(123456)\ep(1234)}\kappa+\frac{s_{13}s_{32}s_{24}(1|3+2|4)^2}{\ep(132456)\ep(1324)}\kappa+\frac{s_{23}s_{45}s_{56}s_{61}(4|5+6|1)^2\ep(1234)}{\ep(234561)\ep(324561)\ep(4561)}\kappa,
\end{equation}
while on the box integral side, its coefficient (times $s_{23}$) contains the additional term
\begin{equation}\label{a:boxintegralBCJ1match}
  -\frac{s_{14}(1|2+3|4)^2}{\ep(1234)\ep(4561)}\kappa\cdot s_{23}.
\end{equation}
We have verified numerically that \eqref{a:boxintegralBCJ1new} and \eqref{a:boxintegralBCJ1match} are indeed identical.

For the type involving $I_4(ab|cd|e|f)$ (e.g.~$I_4(12|34|5|6)$), the pentagon integrals $I_5(1|2|34|5|6)$ and $I_5(2|1|34|5|6)$ together with the constraint give additional terms in \eqref{a:boxintegralBCJ2}
\begin{equation}\label{a:boxintegralBCJ2new}
  -\frac{s_{56}s_{61}s_{12}(5|6+1|2)^2}{\ep(123456)\ep(5612)}\kappa+\frac{s_{56}s_{62}s_{21}(5|6+2|1)^2}{\ep(213456)\ep(5621)}\kappa-\frac{s_{12}s_{34}s_{45}s_{56}(3|4+5|6)^2\ep(5612)}{\ep(123456)\ep(213456)\ep(3456)}\kappa.
\end{equation}
And numerically, we have also checked that this is equivalent to the additional term given by the corresponding box integral (times $s_{12}$)
\begin{equation}\label{a:boxintegralBCJ2match}
  \frac{s_{56}(5|3+4|6)^2}{\ep(3456)\ep(5612)}\kappa\cdot s_{12}.
\end{equation}

For the remaining type involving $I_4(abc|d|e|f)$ (e.g.~$I_4(123|4|5|6)$), we first analyze the relation from $I_5(1|23|4|5|^)$ and $I_5(23|1|4|5|6)$, which together with the constraint adds into \eqref{a:boxintegralBCJ3part1} the following terms
\begin{equation}\label{a:boxintegralBCJ3part1new}
  \begin{split}
    &-\frac{s_{45}s_{56}s_{61}(4|5+6|1)^2}{\ep(234561)\ep(4561)}\kappa+\frac{s_{14}s_{45}s_{56}(1|4+5|6)^2}{\ep(231456)\ep(1456)}\kappa\\
    &+\frac{s_{12}s_{34}s_{45}s_{56}(3|4+5|6)^2\ep(4561)}{\ep(123456)\ep(213456)\ep(3456)}\kappa+\frac{s_{13}s_{45}s_{56}s_{62}(4|5+6|2)^2\ep(4561)}{\ep(134562)\ep(314562)\ep(4562)}\kappa.
  \end{split}
\end{equation}
Then in another relation \eqref{a:boxintegralBCJ3part2}, we would get the following terms
\begin{equation}\label{a:boxintegralBCJ3part2new}
  \begin{split}
    &-\frac{s_{45}s_{56}s_{63}(4|5+6|3)^2}{\ep(124563)\ep(4563)}\kappa+\frac{s_{34}s_{45}s_{56}(3|4+5|6)^2}{\ep(123456)\ep(3456)}\kappa\\
    &+\frac{s_{31}s_{24}s_{45}s_{56}(2|4+5|6)^2\ep(4563)}{\ep(312456)\ep(132456)\ep(2456)}\kappa+\frac{s_{32}s_{45}s_{56}s_{61}(4|5+6|1)^2\ep(4563)}{\ep(324561)\ep(234561)\ep(4561)}\kappa.
  \end{split}
\end{equation}
Similarly, in order for this type of dual kinematics relations to hold, we should expect that \eqref{a:boxintegralBCJ3part1new}$-$\eqref{a:boxintegralBCJ3part2new} matches with the additional terms from the corresponding box integral (times $s_{123}$)
\begin{equation}\label{a:boxintegralBCJ3match}
  \frac{s_{45}s_{56}[46]}{\ep(123456)}\left(\frac{\la34\ra[3|4+5|6\ra}{\ep(3456)}-\frac{\la61\ra\la4|5+6|1]}{\ep(4561)}\right)\kappa\cdot s_{123}.
\end{equation}
And we have also verified this numerically.

\spacesection

\section{Transformation to Explicit Formula for $6$-pt $1$-Loop MHV Gravity Amplitude in Terms of MHV Hexagons}

In this appendix we show that starting from our formula \eqref{eq:gravityamplitude6pt} for $6$-pt $1$-loop MHV amplitude in $\mN=8$ supergravity, which is constructed with MHV hexagons together with the virtual coefficients $\gamma$, it is also possible to obtain a formula, which is expressed explicitly in terms of ordinary kinematics data and meanwhile enjoys correct residues on all contours.

In order to achieve this, pay attention that in general there are many non-planar identities relating different MHV polygons. Specifically for MHV hexagons, a direct analysis on leading singularities shows that, among the $60$ inequivalent MHV hexagons, the maximally independent set only contains $40$, so we would expect a great many identities. The smallest of such identities contains $8$ MHV hexagons, which are related by exchanging labels within each pair, if we regard all the $6$ labels as being separated into $3$ adjacent pairs. One example is
\begin{equation}\label{a:hexagonidentity}
  \begin{split}
    \mP_6(123456)-\mP_6(123465)-\mP_6(124356)+\mP_6(124365)&\\
    -\mP_6(126543)+\mP_6(125643)+\mP_6(126534)-\mP_6(125634)&=0.
  \end{split}
\end{equation}
Since these are purely mathematical relations, we are allowed to freely add them with any coefficient into the amplitude and meanwhile preserve all necessary physical data.

Now go back to our formula for $6$-pt gravity amplitude. The trick is to modify the coefficient of each term appearing in \eqref{eq:gravityamplitude6pt}, say for $\gamma(123456)$
\begin{equation}\label{eq:coefficienttrick}
  \frac{\gamma(123456)}{\la123456\ra^2}=-\frac{[16][23][45]}{\la16\ra\la23\ra\la45\ra}\frac{\gamma(612345)}{\la612345\ra\ep(612345)}-\frac{[12][34][56]}{\la12\ra\la34\ra\la56\ra}\frac{\gamma(123456)}{\la123456\ra\ep(123456)},
\end{equation}
and the rest can be obtained just by permutations. A straightforward combinatoric counting gives $60$ inequivalent MHV hexagons, so after the modification we have doubled them to $120$. Then notice that now in front of each term, we always have a pre-factor of the pattern
\begin{equation*}
  \frac{[ab][cd][ef]}{\la ab\ra\la cd\ra\la ef\ra}.
\end{equation*}
And in the $6$-pt case there are altogether $15$ of them. Then it is easy for us to check that the $120$ terms in the modified formula exactly fall into $15$ groups, each involving $8$ terms, and with the same pre-factor. If we denote
\begin{equation*}
  \widetilde{\gamma}(abcdef)=\frac{\gamma(abcdef)}{\la abcdef\ra\ep(abcdef)},
\end{equation*}
then for example, one of the group is
\begin{equation}
  \begin{split}
    -\frac{[12][34][56]}{\la12\ra\la34\ra\la56\ra}\big[
    &\widetilde{\gamma}(123456)\mathcal{P}_{123456}+\widetilde{\gamma}(123465)\mathcal{P}_{123465}+\widetilde{\gamma}(124356)\mathcal{P}_{124356}+\widetilde{\gamma}(124365)\mathcal{P}_{124365}\\
    +&\widetilde{\gamma}(125634)\mathcal{P}_{125634}+\widetilde{\gamma}(125643)\mathcal{P}_{125643}+\widetilde{\gamma}(126534)\mathcal{P}_{126534}+\widetilde{\gamma}(126543)\mathcal{P}_{126543}\big].
  \end{split}
\end{equation}
But again, all $\widetilde{\gamma}$ are just simply related by the constraint \eqref{eq:gammaconstraint} (since here we are only interested in obtaining an explicit example, for simplicity we just set $\kappa=0$). For this group we can bring the other $\widetilde{\gamma}$ to $\widetilde{\gamma}(123456)$ with some additional terms purely of ordinary spinors, and when we collect all of the coefficient of the remaining $\widetilde{\gamma}(123456)$, we get
\begin{equation}
  \begin{split}
    -\frac{[12][34][56]}{\la12\ra\la34\ra\la56\ra}\widetilde{\gamma}(123456)\big[&\mathcal{P}_{123456}-\mathcal{P}_{123465}-\mathcal{P}_{124356}+\mathcal{P}_{124365}\\
    -&\mathcal{P}_{125634}+\mathcal{P}_{125643}+\mathcal{P}_{126534}-\mathcal{P}_{126543}\big].
  \end{split}
\end{equation}
However, the linear combination of the eight MHV hexagons in the big square bracket is just one of the standard identities of these kind of mathematical objects as shown in \eqref{a:hexagonidentity} that should always vanish. Since the same situation can be shown to occur in all other groups, to this point we have shown the $\gamma$ factors are completely eliminated from the entire formula, and what is left over is just the $1$-loop gravity amplitude in the ordinary spin-helicity formalism, despite the fact that now the expression would look quite complicated and asymmetric.

\acknowledgments

E.Y.Y.~would like to give thanks to Freddy Cachazo for specifying this problem and supervising over this work during E.Y.Y.'s stay in the Perimeter Scholars International program, as well as Andrew Hodges, Song He, Sayeh Rajabi and Stephen Naculich for useful discussions. This work is supported by the Perimeter Institute for Theoretical Physics. Research at Perimeter Institute is supported by the Government of Canada through Industry Canada and by the Province of Ontario through the Ministry of Research \& Innovation. E.Y.Y.~is supported in part by the NSERC of Canada and MEDT of Ontario.


\bibliographystyle{JHEP}
\bibliography{VirtualCKDuality}




\end{document}